# Effect of cations on van der Waals interactions between particles in aqueous alkali nitrate electrolytes

Micah P. Prange[a*], Jaehun Chun[a*], Gregory K. Schenter[a], Elias Nakouzi[a], Yihui Wei[b], Aurora E. Clark[b], Kevin M. Rosso[a], Carolyn I. Pearce[a]

[a] Pacific Northwest National Laboratory, Richland, Washington 99354, United States

[b] Department of Chemistry, University of Utah, Salt Lake City, Utah 84112, United States

## Abstract

The van der Waals interaction has been extensively studied for colloidal forces and resultant emergent phenomena such as colloidal stability, aggregation, and suspension rheology, but the effect of electrolytes on this interaction, especially at intermediate and high electrolyte concentrations, remains incompletely understood. We have extended the Lifshitz theory for van der Waals interactions in pure water to alkali nitrate solutions at arbitrary concentrations by developing a dielectric response model for alkali nitrate solutions that is based on electronic structure calculations of the molecular constituents. Due to their importance in catalysis, ceramics, and coating technologies, the Hamaker constants for rutile, boehmite, and alumina nanoparticles suspended in alkali nitrate solutions are calculated as a function of salt concentration. Contrary to prevailing assumptions, increasing the concentration of sodium (Na), potassium (K), and rubidium (Rb) nitrate solutions causes appreciable increases of the Hamaker constants relative to pure water instead of decreases, whereas cesium nitrate ($CsNO_3$) has almost no effect on the Hamaker constant. We discussed the influence of the solution molar volume, the polarizability of the dissolved ions, and optical properties of the interacting particles in the context of previously published work. Our study indicates a non-vanishing role of van der Waals interactions on colloidal stability at intermediate and high electrolyte concentrations, leading to physical insights on emergent phenomena associated with nanoparticles.

## 1. Introduction

The attractive van der Waals interaction is ubiquitous in a wide range of physicochemical and biological problems such as adsorption of molecules/macroscopic entities, crystal growth, cloud microphysics, nanoparticle self-assembly, rheological responses of colloidal suspensions, DNA condensation, and protein folding.[1-4] Therefore, a fundamental understanding of van der Waals interactions in various chemical environments is critical to obtain physical insights into such natural and man-made processes.

A modern theory of van der Waals interactions, called the macroscopic or continuum theory, was developed by Lifshitz[5] and Dzyaloshinski et al.[6] In contrast to Hamaker's previous theory based on a pairwise summation of the London potential over the volumes of two interacting entities, the modern theory included many-body effects by placing the van der Waals interaction in the context of quantum electromagnetic modes in analogy to Planck's approach for blackbody radiation[7] (i.e., correlated behavior arising from interfering electromagnetic waves emanating from individual constituents such as particles and the intervening medium). However, a full-fledged Lifshitz formulation has been evaluated only for limited geometries, such as semi-infinite flat surfaces[6, 8] and spherical particles.[9] Therefore, an "effective" Hamaker constant obtained from the Lifshitz theory based on semi-infinite flat surfaces (and often employing simplifying mathematical assumptions such as Taylor series expansion of the wave vector integral and neglect of electromagnetic retardation)[8, 10] is commonly used to calculate the van der Waals interaction as a function of geometrical parameters such as size and separation, sometimes including mutual orientation for non-spherical particles.[11]

The most fundamental data required for the calculation of the effective Hamaker constant are the dielectric functions of the materials composing the individual entities involved over a large range of frequencies (from zero to exahertz). Most previous work has focused on the frequency-dependent dielectric response of organic materials (e.g., hydrocarbons and polymers)[12, 13] and amorphous or crystalline inorganic materials (e.g., silica, mica, titania),[14] providing important physical insights on attractive interactions between macroscopic entities in various aqueous environments. Typically, the dielectric function is represented via a Lorentzian-type oscillator model with several characteristic frequencies, oscillator strengths, and damping parameters based on reasonable fitting to available experimental and computational data.

However, the inclusion of such responses of the intervening aqueous medium has been limited to pure water in practice, although static dielectric constants and dielectric functions at low-frequency (up to $O(1)$ THz) of aqueous solutions up to intermediate/high concentrations of electrolytes have been studied,[15-17] and ion-specific effects were observed. The dielectric response in the UV-visible frequency range can have a large effect on the Hamaker constant due to the large number of Matsubara "sampling" frequencies contributing (c.f., Eq. (1)). This poses a significant challenge to understanding the role of electrolyte concentrations on van der Waals interactions up to high concentrations of electrolytes (i.e., $O(1)$ m concentration), because obtaining accurate knowledge of the dielectric response at high frequencies is difficult. Hence, fundamental understanding of colloidal stability at higher electrolyte concentrations and resultant rheological responses has not been developed. This knowledge gap is crucial for multiple applications involving extreme chemical environments, such as treatment of industrial runoff, acid mine drainage, and processing of legacy nuclear waste at the Department of Energy's former plutonium production site, Hanford[18].

Electrolytes have been included in the Lifshitz theory as charge fluctuations.[8] Following this conceptual framework, the screening due to ions plays an analogous role to retardation screening by finite-frequency electromagnetic fluctuations in the original Lifshitz theory. This formulation involves Debye-Hückel theory for the Poisson-Boltzmann equation, which predicts an exponential screening factor, $e^{-\kappa L}$, where $\kappa^{-1}$ is the Debye screening length and $L$ is a characteristic separation distance between objects of interest.[8, 19, 20] Besides an assumption of ideal solution behavior that limits its applicability to dilute conditions, this treatment neglects the frequency dependence of the change to the dielectric response due to the presence of electrolytes. As a result, the van der Waals interaction is always suppressed by electrolytes in this formulation. Furthermore, the validity of this framework becomes questionable at intermediate and high electrolyte concentrations when the concept of the Debye length becomes poorly defined in regimes of underscreening and associated Kirkwood transitions.[21, 22]

In this work, we construct a model that predicts the dielectric responses of aqueous alkali nitrate solutions ($MNO_3$, where M=Na, K, Cs, Rb) over a range of concentrations by extending foundational work on pure water.[13] We then use this model to compute the compositional dependence of the Hamaker constant describing van der Waals interactions between nanoparticles of rutile, boehmite, and alumina in aqueous electrolyte solutions, as these are key mineral phases in catalysis, ceramics, coating technologies, and nuclear waste processing. We compare our work to existing results in the literature and present recommendations for obtaining Hamaker constants for force calculations in concentrated solutions, followed by a summary of our findings and their implications.

## 2. Methods

The main quantity of interest is the nonretarded effective Hamaker constant:[23]

$$A_H(T) = \frac{3}{2}k_{\mathrm{B}}T\sum_n{}' \int_0^\infty dx\, x\ln(1-\Delta_n^2 e^{-x}) = \frac{3}{2}k_{\mathrm{B}}T\sum_n{}'\sum_{m=1}^{\infty}\frac{\Delta_n^{2m}}{m^3}. \tag{1}$$

Here, $\Delta_n^2 = \left(\frac{\varepsilon_A(i\xi_n)-\varepsilon_m(i\xi_n)}{\varepsilon_A(i\xi_n)+\varepsilon_m(i\xi_n)}\right)^2$ depends on the dielectric functions of the material composing the particles ($\varepsilon_A$) and the solution ($\varepsilon_m$) evaluated on the imaginary axis of the complex frequency plane at the $n^{\mathrm{th}}$ Bosonic Matsubara frequency $i\xi_n = 2\pi i n k_{\mathrm{B}}T/\hbar$, called the van der Waals spectra of individual entities. The constants $k_{\mathrm{B}}$ and $\hbar$ are the Boltzmann constant and the reduced Planck constant, respectively, and $\sum_n' = {}^1\!/_2\sum_{n=0}^{0} + \sum_{n=1}^{\infty}$ . The focus of this work is to formulate models of the dielectric response of alkali nitrate solutions ($\varepsilon_m$) and minerals ($\varepsilon_A$) that allow the evaluation of Eq. (1).

### 2.1 Model of dielectric response for aqueous $NaNO_3$ solutions

The solutions were modeled as collections of independent molecular and ionic species: $H_2O$, $NO_3^-$, and $M^+$. Each of these species was assigned a frequency-dependent molecular/ionic oscillator strength distribution, and the total dielectric spectrum of the solution was assembled by linear combination of the molecular contributions in proportion to their population in the solution. The form of the dielectric function of the solution is the same as used by Parsegian and Weiss[13] (PW hereafter):

$$\epsilon(\omega) - 1 = \sum_j \frac{\tilde{f}_j - i\omega h_j}{\omega_j^2 - \omega^2 - i\omega g_j}. \tag{2}$$

This form enforces the analytic structure of the dielectric function and can be evaluated at any complex frequency except the isolated poles where the denominator vanishes. In this work, the sum over $j$ runs over oscillators associated with the solvent and solute species. Each species hosts several oscillators that numerically describe the electronic transitions arising from its occupied orbitals. We considered only electronic transitions in the visible and UV regions and neglect Debye-type relaxation ($h_j = 0$ , a limitation to revisit in future work) and put the oscillator strength in molecular (or molar) terms: $\tilde{f}_j = n_j f_j$ where $n_j$ is the number density (or inverse molar volume) of the molecular/ionic species hosting the $j^{\mathrm{th}}$ oscillator, which has resonance frequency $\omega_j$ and width parameter $g_j$. Since all the excitation energies in this work are more than 5 eV, there is negligible thermal excitation, and the computed dielectric response is calculated at zero temperature. Such an approximation is not appropriate for ionic relaxation.

The dielectric response model can be completely specified by listing: i) the molecular/ionic poles $j$, ii) the mole fractions of the solvent (water) atoms and solute ions, and iii) the density (or molar volume) of the bulk solution. We took the parameters of the oscillator model for molecular water (with the oscillator strengths divided by $n_{\mathrm{H_2O}} \approx 0.0335$ Å$^{-3}$, the density of molecules in liquid water at room temperature) from Ref.[13]. To consistently neglect nonelectronic response, we kept the oscillators with the 6 highest frequencies (out of 11) from the PW model. Note that a qualitative comparison of the PW model with the inelastic X-ray scattering experiment of Hayashi et al.[24] shows good agreement up to about 15 eV.

The transitions of the other electrons (i.e., those belonging to the electrolyte ions) were computed using linear response time-dependent density functional theory (LR-TDDFT)[25] of gas-phase ions. The TDDFT simulations yielded a list of excited states ('roots') characterized by an excitation energy $\omega_j$ and oscillator strength $f_j$ that were used as oscillator descriptors in Eq. (2). In these calculations, we used the PBE0 exchange-correlation functional[26] with aug-cc-pVTZ[27, 28] and def2-TZVPPD[29-31] basis sets for $\mathrm{NO_3^-}$ and the alkali cations, respectively. The associated effective core potentials (ECPs), leaving 8 explicit valence electrons, were applied to $Rb^+$ and $Cs^+$. The Tam-Dancoff approximation was employed, and 50 singlet excitations were found for each alkali metal ion. 200 such excitations were computed for $\mathrm{NO_3^-}$. Reference PBE0-aug-cc-pVTZ simulations were also performed for the gas phase water molecule, even though the PW pole model was used to represent the dielectric response of water in the results. The PW dielectric function is compared to its TDDFT counterpart in Fig. 1.

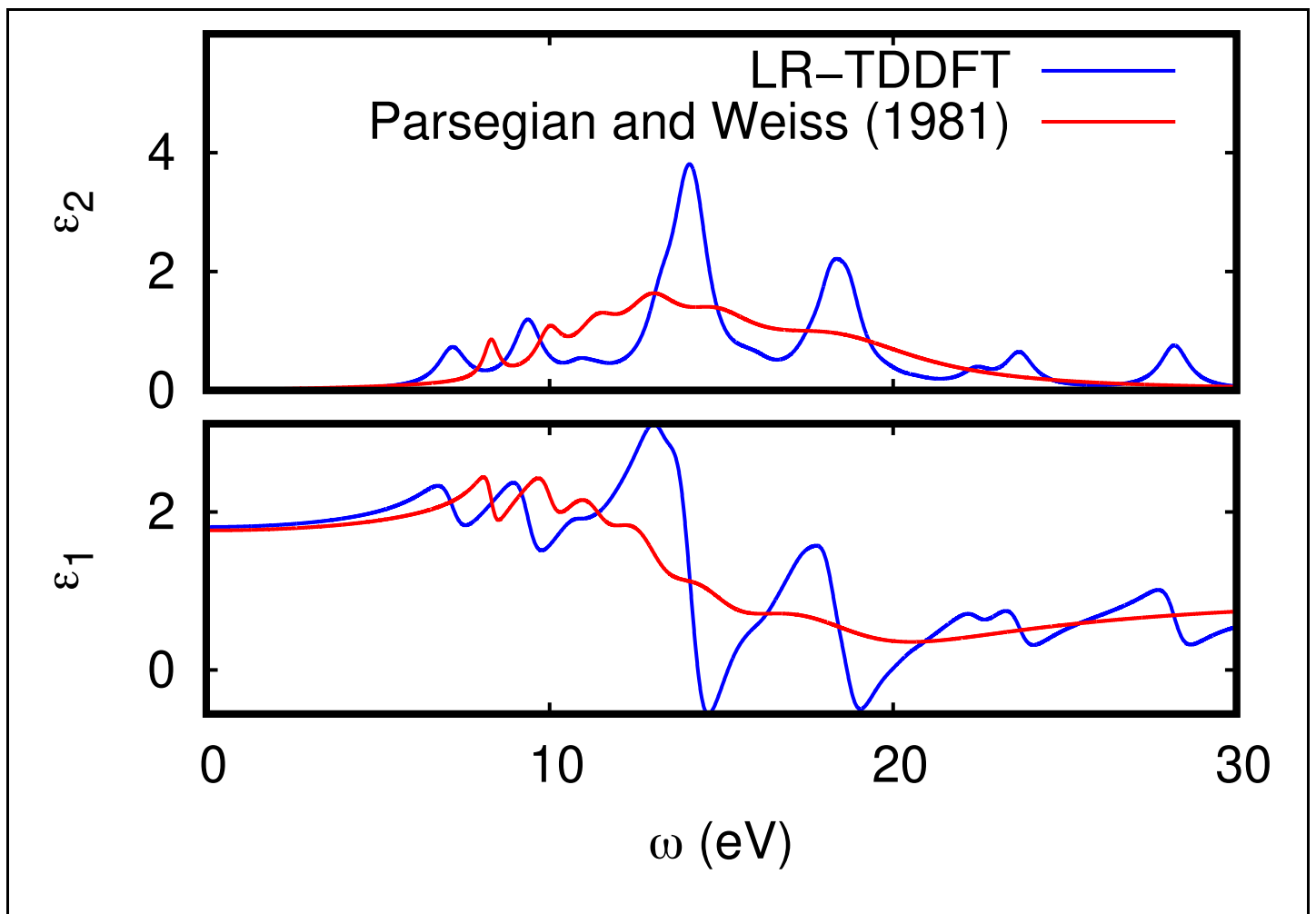

**Figure 1:** Real (bottom) and imaginary (top) parts of the dielectric function of water using PBE0-aug-cc-pVTZ TDDFT simulations (blue) and the pole model of Reference [13] (red).

In addition to the molecular/ionic responses, the concentrations of the different species are needed to evaluate Eq. (2) (and subsequently Eq. (1)) to compute the Hamaker constant. In this work we used curves fitted to the concentration dependence of the apparent molal volume $\phi_v$ of the salt from Berchiesi et al.[32] of the form

$$\phi_v(c) = \phi_v^0 + S_v\sqrt{c} \quad (3)$$

where $c$ is the molarity of the solution and $\phi_v^0$ and $S_v$ are fit parameters. Although the fits were done at low concentrations, the values predicted by Eq. (3) agree with theoretical estimates based on molecular dynamics simulations of aqueous $NaNO_3$ solutions[33] to better than 10% at 8.7 m. The derived number densities, which are needed to evaluate Eq. (2), are plotted in Fig. 2, which also includes direct molecular dynamics determinations for $NaNO_3$ (black '+'s). Note that the theoretically determined densities are *lower* at high concentrations than the model of Berchiesi et al. despite the prevalence of ion clustering in the molecular dynamics simulations.[33] The volume of the alkali metal cation grows with atomic number leading to a steeper decrease of the water number density with concentration for larger alkali metals, which, as discussed below, is consequential for the Hamaker constant calculations. This behavior is consistent with the trends in the ionic radius of the hydrated alkali metal cations. Relevant ionic radii as determined by Mahler and Persson[34] are collected in Table 1.

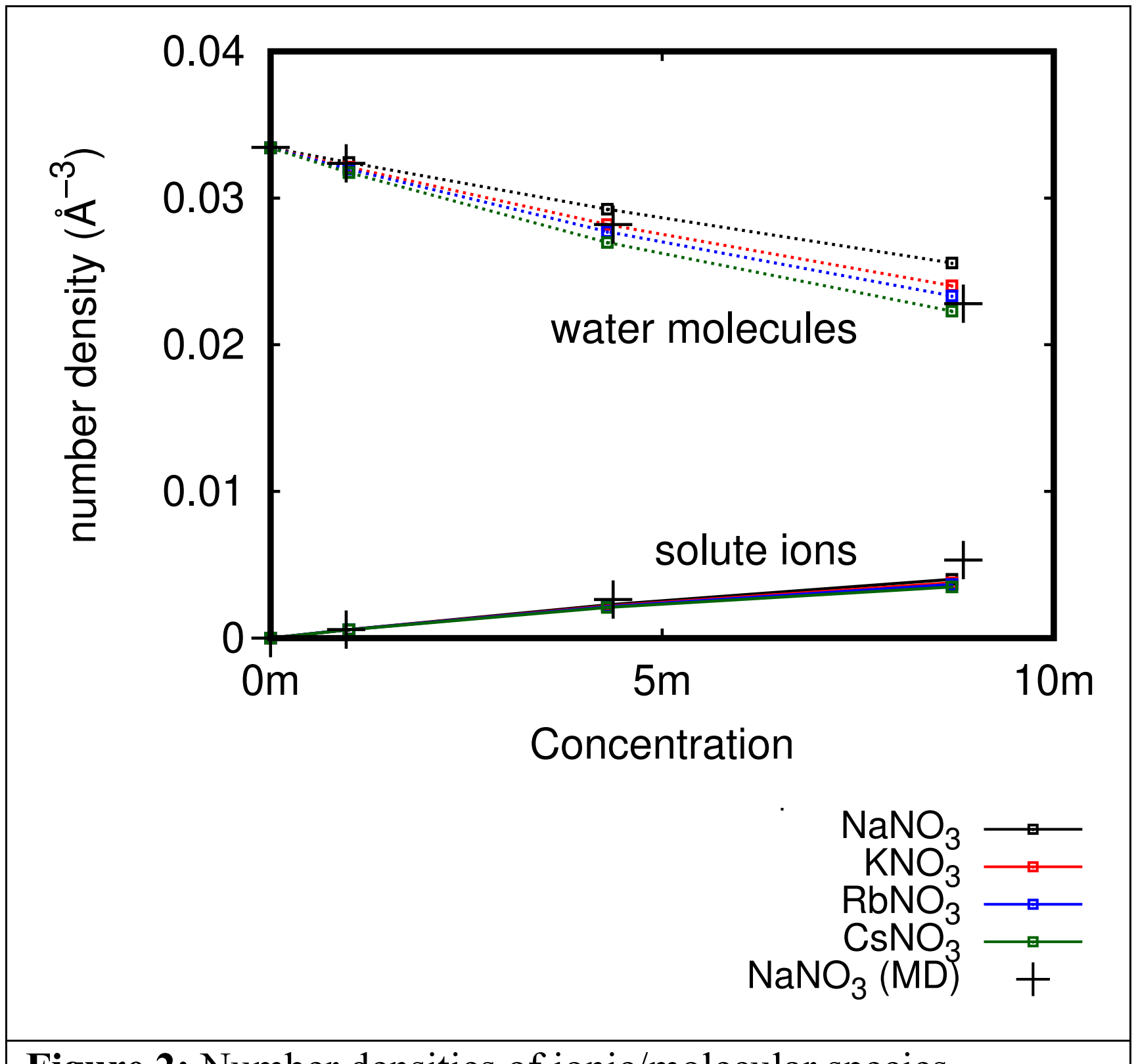

**Figure 2:** Number densities of ionic/molecular species predicted by the fits of Ref. [32] (lines) and simulations of Ref. [33] ('+'s).

**2.2 Model of dielectric response for rutile, boehmite, and alumina**

We calculated the dielectric functions for the bulk mineral phases considered here using plane-wave TDDFT,[35] extending our previous work on boehmite and alumina[36] to rutile. For consistency, the same hybrid exchange-correlation functional (PBE0) was used as in the local orbital calculations of the ions described above. The projector-augmented wave (PAW) method was used to represent core electrons leaving 6, 4, 3, and 1 explicit valence electrons for O, Ti, Al, and H, respectively.[37] The core electrons are not present in the simulations, so their excitations do not contribute to the results reported here. The first neglected transitions are the Al $L_{23}$ edges, near 70 eV. They could be included following the methods of Kas et al.,[38] but their effects on the dielectric function are small as can be seen for the case of alumina in the data of Hagemann et al.[39] Similar comments apply to the Cs and Rb ions. Starting structures describing rutile in the $P4_2/nnm$ space group[40] and $\alpha$-alumina (sapphire) in $R\bar{3}c$[40, 41] (i.e., the corundum rhombohedral cell) were obtained from the literature. The plane wave basis was determined by an energy cutoff of 500 eV and 6x2x6 (boehmite), 3x3x5 (rutile), and 4x4x4 (alumina) k-point grids. The computed band gaps (direct at Γ) are 1.67, 5.23, and 6.27 (PBE) and 3.87 eV, 8.12, and 8.96 eV (PBE0) for rutile, boehmite, and alumina, respectively. These results are in general agreement with previous calculations.[36, 42, 43] The van der Waals spectrum is then calculated by Kramer-Kronig transformation of the computed imaginary part $\epsilon_2$ of the dielectric function using the expression[13]

$$\epsilon(i\xi) - 1 = \frac{2}{\pi}\int_0^\infty d\omega \frac{\omega\epsilon_2}{\omega^2 + \xi^2}. \tag{4}$$

All three minerals considered here have anisotropic (uniaxial) optical properties. We used spherical averages to compute the Hamaker constants reported here, consistent with an assumption of random mutual orientation of particles in the suspension. In this work, we have chosen to use consistent approximate representations of the dielectric response of all contributing species to provide a framework to estimate trends with solution composition. The accuracy of all the dielectric functions could be improved, which is a topic of our future research.

## 3. Results and Discussions

### 3.1 Electronic structure and UV response of ionic constituents

The absorption spectra of the molecular constituents of the solutions that are used in our Hamaker constant calculations are plotted in Fig. 3. Specific details on the absorption spectra for anions and cations are explained below:

**Nitrate anion:** The absorption spectrum of the aqueous $NO_3^-$ consists of bound-to-bound intramolecular transitions, dominated by strong π-π* transitions, and bound-to-continuum transitions at higher frequencies. Harris calculated the dipole allowed π-π* transition of nitrate to occur at 6.5 eV with an oscillator strength of ~0.3 using CNDO methods.[44] Pedersen et al.[45] located the same transition at 6.75 eV with a higher oscillator strength (~0.4 for gas phase and 0.34 in aqueous solution) using TDDFT. Our gas phase calculations place the transition at 6.42 eV with oscillator strength 0.35, the eighth excitation of 200 that were calculated. The resulting photo-absorption cross-section spectrum is plotted as the blue curve in Fig. 3.

**Alkali cations:** These ions have rare gas configurations in which $np$ with $n$=2, 3, 4, 5 for $Na^+$, $K^+$, $Rb^+$, $Cs^+$, respectively is the highest occupied orbital. The absorption spectrum is dominated by two sets of strong, dipole allowed transitions from $np$ to $(n+1)s$ and $(n+1)d$ orbitals. The increasing principal quantum number of the optically active orbitals leads to decreasing transition energies and more intense transitions with increasing atomic number. The energies and strengths of these transitions are collected in Table 1. This trend affects the contribution of the alkali metal ions to the integrand in Eq. (1), resulting in a noticeable cation effect in the computed Hamaker constant as discussed below.

**Table 1:** Plausible aqueous coordination numbers and ionic radii[34] and UV transition energies and oscillator strengths for the dominant transitions for the alkali metal cations.

| | **Aqueous solution** | | **np-(n+1)s** | | **np-(n+1)d** | |
|---|---|---|---|---|---|---|
| | Coordination number /ionic radius (Å) | | Frequency (eV) | Oscillator strength | Frequency (eV) | Oscillator strength |
| $Na^+$ | 5/1.02 | 6/1.07 | 31.2 | 0.28 | 38.9 | 0.71 |
| $K^+$ | 6/1.38 | 7/1.46 | 19.8 | 0.67 | 24.2 | 2.4 |
| $Rb^+$ | 8/1.64 | | 16.5 | 0.84 | 20.4 | 2.0 |
| $Cs^+$ | 8/1.73 | | 13.8 | 1.0 | 16.9 | 4.7 |

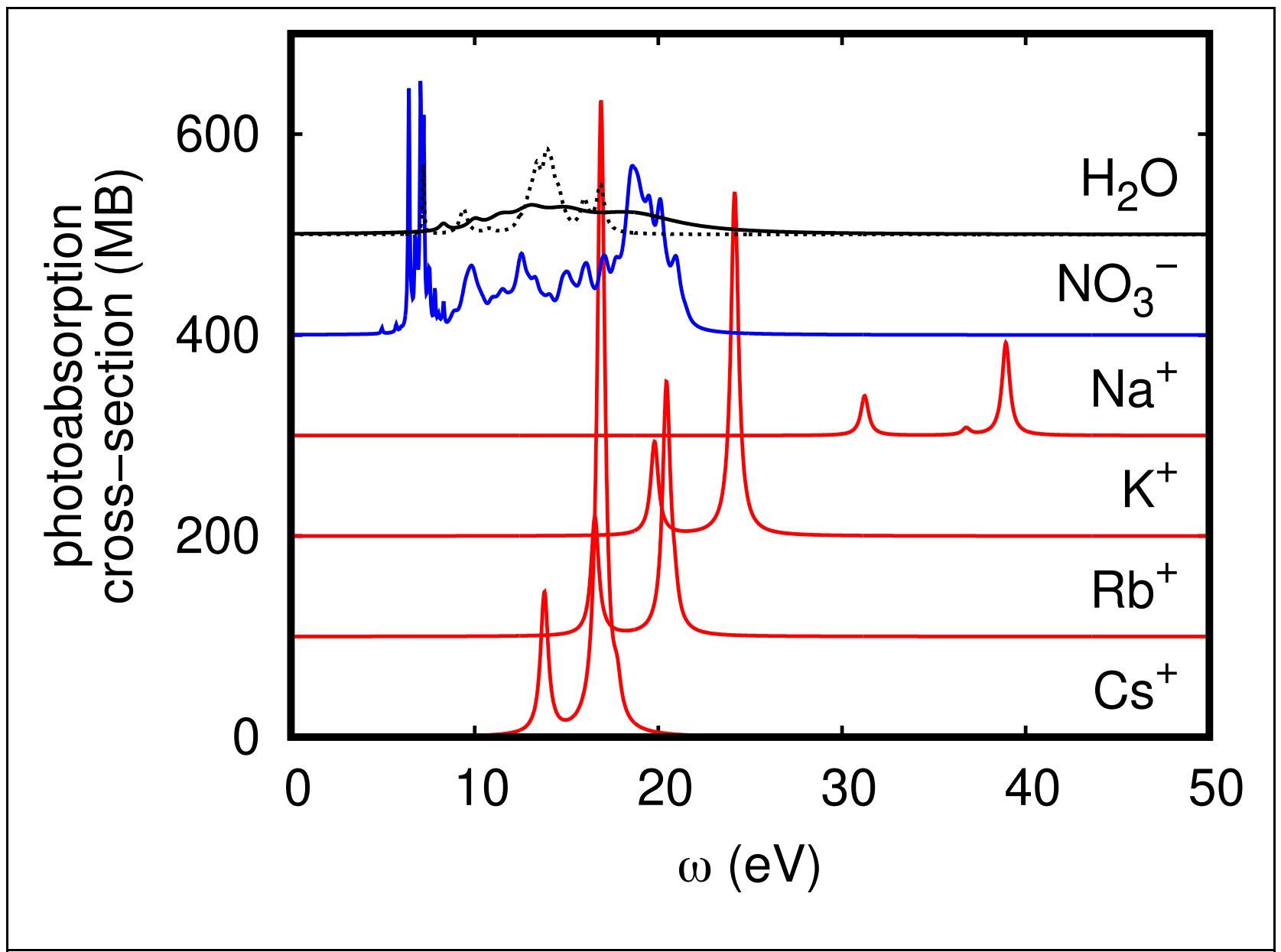

**Figure 3:** Photoabsorption cross-section distributions for water in the PW model (soid black) and TDDFT (dotted black) and $NO_3^-$ (blue) and alkali metal cations (red). Curves offset vertically for clarity.

**3.2 Variation of the dielectric function and Hamaker constants with concentration**

Figure 4 plots the dielectric functions for 8.7 m solutions of the different $NO_3^-$ salts (panels a and b) and boehmite/rutile/alumina (panels c and d), with the imaginary part on the real axis and the real part on the imaginary axis, i.e., the van der Waals spectrum. The spectrum for pure water with no additional ions is also shown in all panels of Figure 4 for reference. The van der Waals spectra of the solution and the particle materials is sufficient to evaluate Eq. (1). The Hamaker constant as a function of the cation species and ionic concentration, found by evaluating Eq. (1) using the model dielectric function described above, is plotted in Fig. 5. For pure water (i.e., the left side of the plots in Fig. 5) the current work differs from previously reported values. We compare them below, summarizing values in Table 2. As the concentration of the solutions is increased, the computed Hamaker constants vary weakly by less than 25% at extreme concentrations up to 10m. In general, the Hamaker constant increases with concentration. This weak increase of $A_H$ with electrolyte concentration is a primary conclusion of this work, opposed to the prevailing assumptions of decreasing Hamaker constant with increasing electrolyte concentration as, e.g., in our previous work.[46]

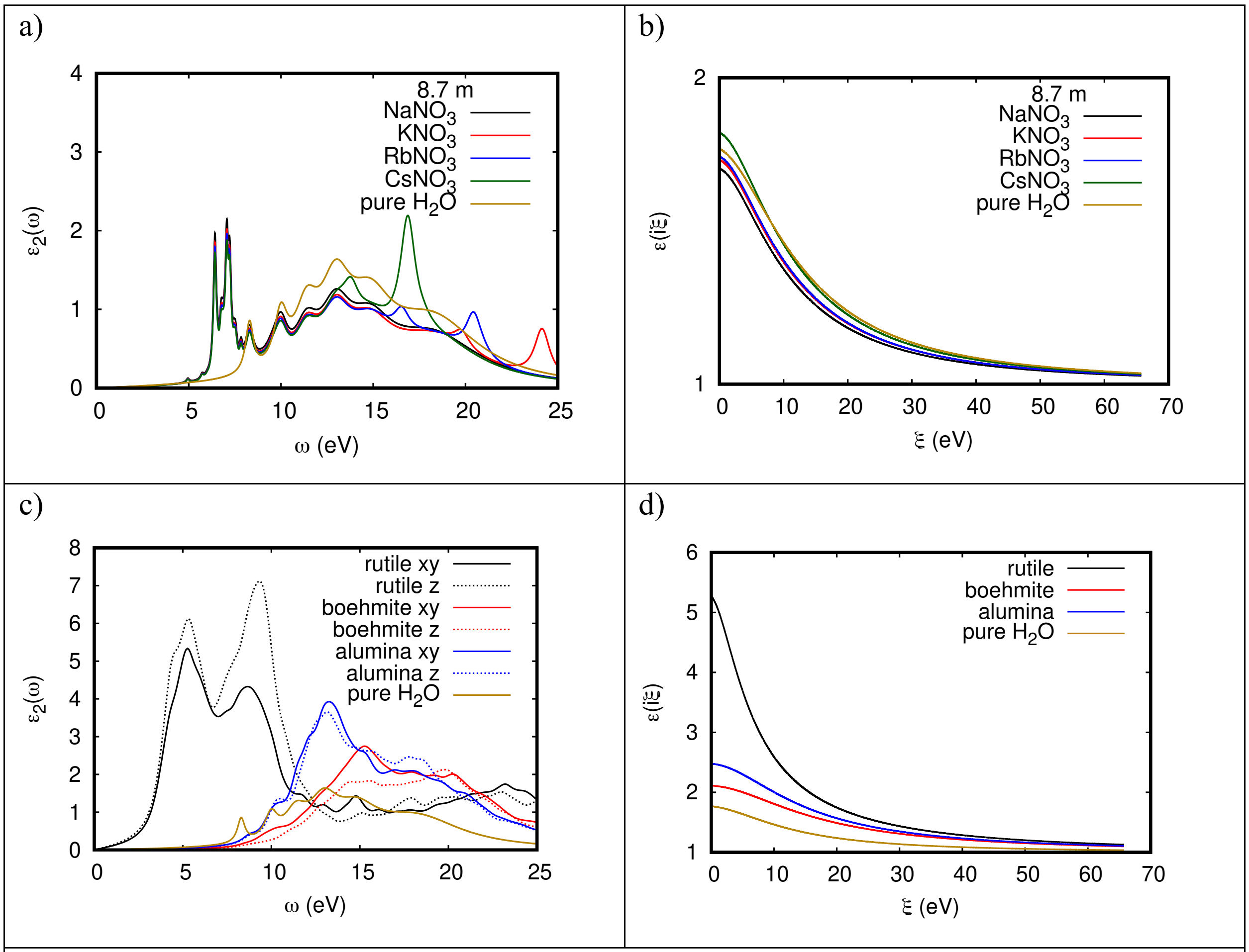

**Figure 4:** Imaginary part of the dielectric function $\varepsilon_2(\omega)$ (a,c) and van der Waals spectrum $\varepsilon(i\xi)$ (b,d) for 8.7M aqueous nitrate solutions with various alkali metal cations (a,b), for crystalline rutile, boehmite, and alumina (c,d), and for pure water (all panels). The spectra of the salt solutions were evaluated within the dielectric response model described in the text, the mineral phases using plane-wave TDDFT, and the water spectrum using the PW[13] model.

Due to the higher spatial density of electronic transitions in the particle materials (rutile, boehmite, and alumina) compared to the aqueous solutions, the van der Waals spectrum of the minerals is greater than all the salt solutions at all frequencies (Fig. 4b,d). Hence the term $\Delta_n^2 = \left(\frac{\varepsilon_A(i\xi_n)-\varepsilon_m(i\xi_n)}{\varepsilon_A(i\xi_n)+\varepsilon_m(i\xi_n)}\right)^2$ appearing in Eq. (1) is a decreasing function of $\varepsilon_m(i\xi_n)$, the van der Waals spectrum of the solution evaluated at the $n^{\text{th}}$ Matsubara frequency, for all $n$. A consequence of this nature is that changes in the solution composition that increase the oscillator strength density of the solution in the UV-vis region decrease the Hamaker constant and vice versa. The electronic resonances situated in the space between two dispersed crystallites tend to screen the long-range van der Waals force between the particles. As the concentration of the solution is increased, there are two main physical effects: i) the volume per mole of water increases to accommodate the dissolved ions, and ii) electronic transitions corresponding to excitation of electrons belonging to the ionic species are introduced. These two effects tend to move $\varepsilon_m$ in opposite directions and hence mutually cancel; the expansion of the volume decreases $\varepsilon_m$, but the "new space" is filled with new electronic resonances that increase $\varepsilon_m$. The direction of the net effect (i.e. whether $A_H$ and the strength of the van der Waals attraction become larger or smaller) is very subtle; it is determined by which trend is more influential in the sum over Matsubara frequencies in Eq. (1). For the cases considered here, at $T$=300K, the effect of volume expansion prevails. As the concentration increases, $A_H$ grows. For Na, K, and Rb nitrate solutions, there is a modest increase in $A_H$. For $CsNO_3$, however, there is almost no concentration effect on $A_H$. The large polarizability of $Cs^+$ at physically relevant frequencies below 20 eV effectively counteracts the volume increase associated with dissolving the ions. Although the effects are small, there is a reversal as the alkali metals are traversed with $A_H$ enhancement in the order $\mathrm{Rb} > \mathrm{Na} \approx \mathrm{K} > \mathrm{Cs} \approx 0$. This ordering is not an artifact of the use of ECPs to neglect core polarization for Rb and Cs, since these cations bracket those for which all electrons were considered. The ordering is the same for rutile, boehmite, and alumina particles interacting across nitrate solutions. Ion clustering at high concentration, which is neglected in the estimation of the solution density, might be expected to counter the volume enhancement, but the available molecular dynamics simulations show that, at least in the case of $NaNO_3$, the current treatment underestimates increase in volume with concentration as mentioned above.

To illustrate the directions of the two countervailing effects we have plotted an auxiliary result obtained by neglecting the $f_j$ in Eq.(2) originated from $Cs^+$ in the $CsNO_3$ calculations in broken green lines in Figure 5 ("sterile Cs"). This unphysically charged solution shows significant enhancement of $A_H$.

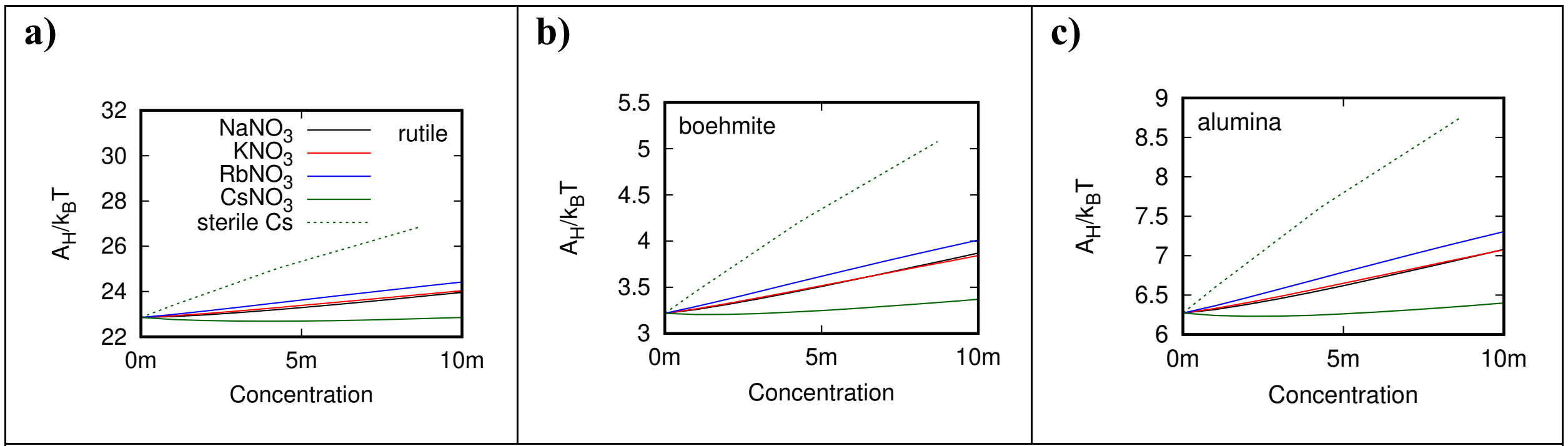

**Figure 5:** Computed effective Hamaker constant at $T$ =300 K as a function of salt concentration for (rutile/boehmite/alumina)/$MNO_3$ systems for M=Na (black), K (red), Rb (blue), Cs (green). The dotted green line shows the results of neglecting the $Cs^+$ polarizability in the calculation of $A_H$.

### 3.3 Comparison of current work to previous work using Cauchy plots or theoretical optical constants

As stated, the need to know the dielectric functions of the medium and particle material over a wide frequency range has historically hampered the application of the Lifshitz theory. Hough and White[47] introduced the method of fitting a single pole model of the form Eq. (2), which was a special case of the many pole form introduced by Ninham and Parsegian,[48] using linear regression of "Cauchy plots"[49] of $(n^2 - 1)$ vs. $\omega^2(n^2 - 1)$ made from measurements of the index of fraction $n(\omega)$ in the band gap region. The slope and intercept of the linear fit of the Cauchy plot are related to the location and strength of a single pole that reproduces the long-wavelength behavior of $n$ (and hence $\varepsilon = n^2$). This method provided a practical way to experimentally determine interaction parameters, and Bergström[14] compiled such fits for many inorganic materials in a work that has become a valuable reference for estimation of van der Waals interactions. Hough and White[47] explicitly noted the assumption that the material should possess a 'simple' absorption spectrum to obtain reliable Hamaker constants from Cauchy plots of index of refraction data, but this assumption can generally only be checked if the spectrum is available over a wide frequency range, in which case there would be no need to resort to the Cauchy plot to get the Hamaker constant in the first place. This work provides an opportunity to compare the different approaches quantitatively.

In Fig. 6, we illustrate the situation for rutile and alumina in water by plotting the simulated dielectric functions used in this work along with the single-pole models of Bergström[14] (fit to measured index of refraction data) and a similar models fit to the 0-4 eV range of our calculated data for rutile and alumina. In the case of alumina, spectral data are available over a large frequency range, as reported by Hagemann et al.[39] These data, along with a Cauchy plot fit, have also been included in Fig. 6.

**Table 2:** Nonretarded Hamaker constants $A_H/k_BT$ at room temperature for alumina/water/alumina and rutile/water/rutile interactions using various spectral representations.

| | TDDFT (full) | TDDFT (single pole) | Bergstrom | Bergstrom (single pole) | Hagemann (full) |
|---|---|---|---|---|---|
| $Al_2O_3$ | 6.27 | 5.18 | 8.86 | 10.9 | 10.7 |
| $TiO_2$ | 22.3 | 15.5 | 12.8 | 14.0 | - |

The discrepancy between Bergström[14] and our results, using identical spectral parameters for rutile, arises from our neglect of the first 5 (non-electronic) poles in the PW water model. A consistent theory would include both the low-frequency dielectric response of the particle material and the intervening solution. We point out that the low-frequency behavior of concentrated electrolyte solutions is likely significantly different than the pure water reflected in the PW model. The current work predicts $A_H/k_BT$=3.22 for boehmite/water/boehmite, which is significantly reduced compared to our earlier work[50], in which we reported $A_H/k_BT$=5.2. The difference between the two analyses results from the different boehmite dielectric functions used. The study by Nakouzi et al.[50] used the DFT computed dielectric function reported by Alemi et al.[51] to describe the dielectric response of boehmite, which led to the conclusion that extreme ionic strength can lead to a very slight reduction in particle aggregation rate ascribed to a decrease in Hamaker constant. Our current framework reproduces the same Hamaker constant for boehmite/water/boehmite when using this dielectric function. We chose instead to use the dielectric function from our own work,[36] which updated the study by Alemi et al.[51] to improve the band gap, among other things. It is noteworthy that the calculated dielectric function for rutile, within the TDPBE0 theory used in the current work, overestimates the optical band gap and underestimates the intensity of the absorption just above onset. Hence, we expect the reported $A_H / k_B T$=22.3 for rutile/water/rutile, while a significant upward revision of all previous estimates (e.g., 12.8 by Bergström[14]), still underestimates the true value. It appears that both the Cauchy plot method and PBE0 simulation perform adequately for $Al^{3+}$-containing solids, in which there are no low-lying empty d orbitals, but fail for 3d transition metal oxides, which do not have simple absorption spectra. This comparison clearly demonstrates that a simplified pole model needs to reproduce more than just the index of refraction in the visible region to reliably predict Hamaker constants. The density effects observed in the electrolyte solution dielectric functions suggest that pole models for Hamaker constant calculations could be improved by enforcing constraints based on the density of valence electrons and the optical gap. We plan to develop such constraints in the future.

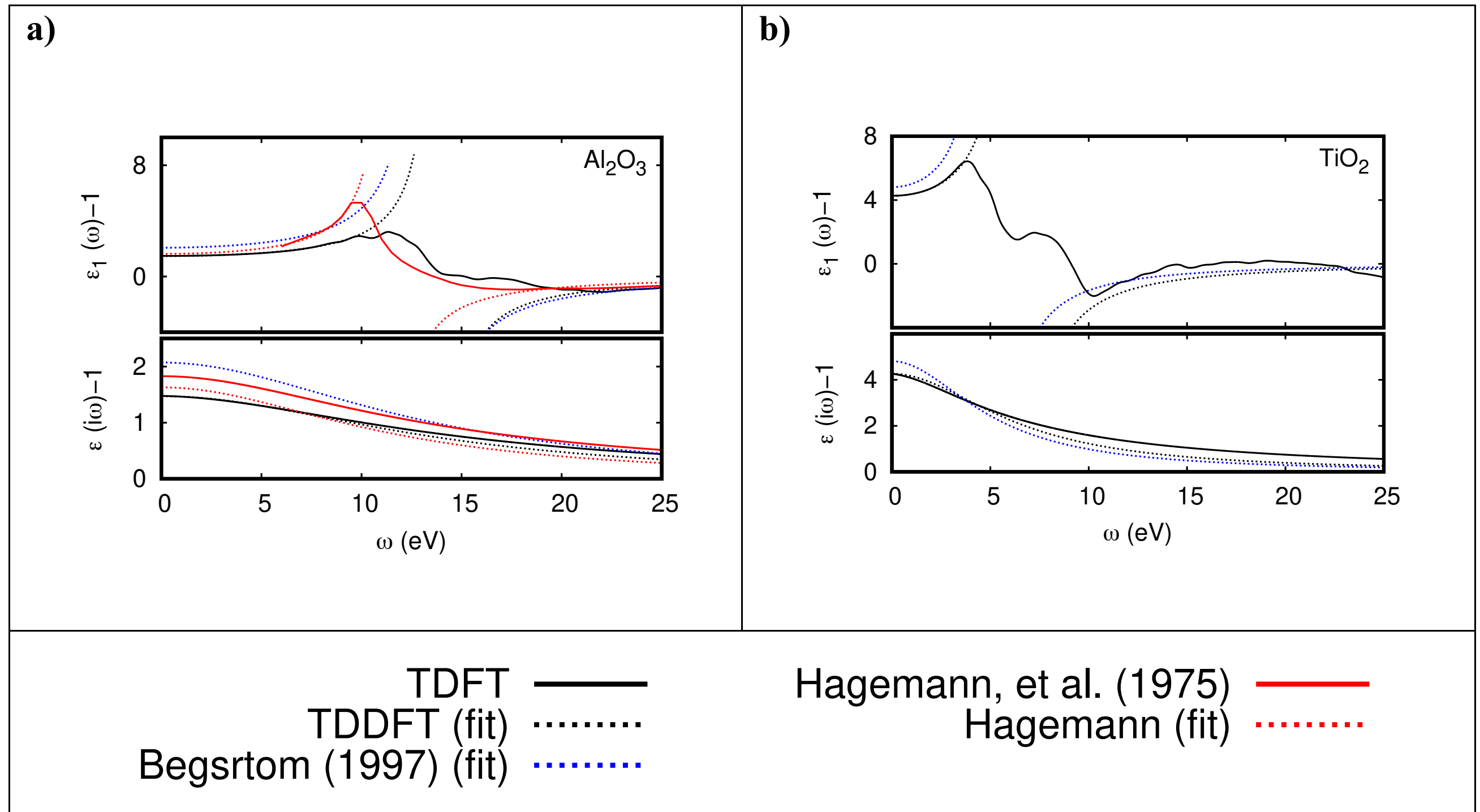

**Figure 6:** The real part of ε-1 of (a) alumina and (b) rutile along the real (top) and imaginary (bottom) axes in frequency space in our TDDFT calculations (black solid line), the single-pole model of Bergström[14] (blue dots), an equivalent single-pole model fit to the TDDFT results (black dots), the measurements of Hagemann et al.[39] (red solid line), and a single-pole fit to the Hagemann data (red dots).

## 4. Conclusions

We have used TDDFT calculations of photo-absoprtion spectra of isolated ions and crystalline nanoparticles to extend the PW treatment of van der Waals interactions in pure water to alkali nitrate solutions over a range of concentrations. This allows estimates of the Hamaker constant by directly calculating the effect of electrolytes on the dielectric response of the intervening medium, and hence the van der Waals forces in the nonretarded limit. This scheme gives a straightforward and practical method for qualitative predictions for the behavior of van der Waals forces in concentrated electrolytes. We find that the Hamaker constants for rutile, boehmite, and alumina interacting across an electrolyte solution grow slowly with concentration in Na, K, and Rb nitrate solutions but are nearly constant in Cs nitrate solutions, in contradiction to previous arguments for an invariable decrease in the Hamaker constant with electrolyte concentration. The concentration effects on the Hamaker constants relevant to suspensions of boehmite and rutile particles in alkali nitrate solutions are small because the changes in the molar volume and the photo-absorption spectrum tend to cancel out in the evaluation of the forces. We further find a sensitive dependence of the computed Hamaker constant on the dielectric function of the particle material.

Particle interactions are critical to understanding aggregation behavior. Considering that the aggregation process is indeed dynamic, the particle interactions influence the aggregation via the relative velocity ($\boldsymbol{U}_r$) between particles based on $\boldsymbol{U}_r = -\boldsymbol{M}_r \cdot \nabla\Psi$ where $\boldsymbol{M}_r$ is the relative hydrodynamic mobility between particles and $\Psi$ is a net particle interaction potential.[52] As noted, the relative velocity that determines a collision time scale between particles directly affects aggregation kinetics. Our study suggests that the effect of the van der Waals interactions on the kinetics of aggregation would be still appreciable at intermediate and high electrolyte concentrations, although physicochemical details such as particle shape and roughness need to be taken into account to quantitatively evaluate the importance of van der Waals interactions.[11] This formalism provides the basis to connect changes in solution structure at the molecular scale to aggregation behavior, which is critical for managing complex slurries in industrial processes, such as the retrieval of legacy nuclear tank waste.

**Data availability**

The data supporting this article are included as part of the ESI.†

## Conflicts of interest

The authors declare no competing interests.

## Acknowledgements

This work was supported by supported by Ion Dynamics in Radioactive Environments and Materials (IDREAM; FWP 68932), an Energy Frontier Research Center funded by the U.S. Department of Energy (DOE), Office of Science, Basic Energy Sciences (BES). Pacific Northwest National Laboratory (PNNL) is a multiprogram national laboratory operated for DOE by Battelle Memorial Institute operating under Contract No. DE AC05-76RL0-1830.